\documentclass{nature}
\usepackage{amsmath,graphicx}
\usepackage{epsf}
\usepackage{epstopdf}
\usepackage{subfigure,epstopdf, textcomp}
\input{epsf}
\usepackage{epsfig}
\usepackage{color}
\usepackage{lscape}
\DeclareGraphicsExtensions{.jpg,.pdf,.png,.eps,.ps}
\usepackage{bm}

\usepackage[paper=letterpaper, top=1.0in, bottom=1in, left=1.0in, right=1.0in]{geometry}

\def\arcsec {\mbox{$^{\prime\prime}$}}
\newcommand{\apj}{Astrophys. J.}

\newcommand{\apjs}{Astrophys. J. Suppl.}

\newcommand{\mnras}{Mon. Not. R. Atron. Soc.}

\newcommand{\aap}{Astron. Astrophys.}

\newcommand{\nar}{New Astronomy Reviews}

\title{Fast Automated Analysis of Strong Gravitational Lenses with Convolutional Neural Networks}

\author{Yashar~D.~Hezaveh\footnote{Hubble Fellow},
Laurence~Perreault~Levasseur,
Philip~J.~Marshall}

\begin{document}

\maketitle

\begin{affiliations}
\item[] Kavli Institute for Particle Astrophysics and Cosmology, Stanford University, Stanford, CA, USA
\item[] SLAC National Accelerator Laboratory, Menlo Park, CA, 94305, USA 
\end{affiliations}

\begin{abstract}
Quantifying image distortions caused by strong gravitational lensing and estimating the corresponding matter distribution in lensing galaxies has been primarily performed by maximum likelihood modeling of observations. This is typically a time and resource-consuming procedure, requiring sophisticated lensing codes, several data preparation steps, and finding the maximum likelihood model parameters in a computationally expensive process with downhill optimizers\cite{Lefor}. Accurate analysis of a single lens can take up to a few weeks and requires the attention of dedicated experts. Tens of thousands of new lenses are expected to be discovered with the upcoming generation of ground and space surveys\cite{LSST,collett}, the analysis of which can be a challenging task. Here we report the use of deep convolutional neural networks to accurately estimate lensing parameters in an extremely fast and automated way, circumventing the difficulties faced by maximum likelihood methods. We also show that lens removal can be made fast and automated using Independent Component Analysis\cite{ICA} of multi-filter imaging data. Our networks can recover the parameters of the Singular Isothermal Ellipsoid density profile\cite{Kormann}, commonly used to model strong lensing systems, with an accuracy comparable to the uncertainties of sophisticated models, but about ten million times faster: 100 systems in approximately $1s$ on a single graphics processing unit.  These networks can provide a way for non-experts to obtain lensing parameter estimates for large samples of data. Our results suggest that neural networks can be a powerful and fast alternative to maximum likelihood procedures commonly used in astrophysics, radically transforming the traditional methods of data reduction and analysis.

\end{abstract}

At its core, lens modeling measures the parameters of a highly nonlinear image distortion. With recent advances in computer vision and deep learning, convolutional neural networks have been shown to excel at many image recognition and classification tasks\cite{ImageNet}. This makes them a particularly promising tool for the analysis of gravitational lenses. Recently, these networks have been used to search for gravitational lenses in large volumes of telescope data\cite{lensfind1,lensfind2,lensfind3} and to simulate weakly lensed galaxy images\cite{lensGAN}. Here we show that these networks can also be used for data analysis and parameter estimation.

We train four networks, Inception-v4\cite{inception}, AlexNet\cite{AlexNet}, Overfeat\cite{overfeat}, and a network of our own design, to analyze strongly lensed systems, by removing their final classification layer and interpreting the outputs of the last fully-connected layer as a prediction for lensing parameters, with all  weights initialized at random. We train the networks to predict the five parameters of the Singular Isothermal Ellipsoid profile: Einstein radius, complex ellipticity, and the coordinates of the center of the lens. We use a squared difference cost function, averaged over the five parameters.
While in many situations in machine learning collecting sufficiently large training sets is one of the main challenges, here we are in the rare situation where it is possible to simulate the training data extremely fast. We train the networks on half a million simulated strong lensing systems. The lensed background sources are composed of three equal sets of images: the first and second comprise real galaxy images from the Galaxy Zoo\cite{GalZoo} machine learning challenge and high-quality images from the GREAT3 training data\cite{GREAT03} and the third set is composed of simulated clumpy galaxies with S\'{e}rsic and Gaussian clump profiles. The position of the background galaxy in the source plane is randomly chosen for each sample but limited to regions where strong lensing occurs, i.e., inside or on the caustics. 

We use a stochastic gradient descent optimizer to train the networks. At each training step, we select a random sample of simulated data, apply randomly-generated, realistic observational effects to each image, and use them to optimize the network weights. 
These effects include convolution with a point spread function (PSF), addition of Poisson shot noise, Gaussian random noise with either a white or colored power spectrum, simulated faint cosmic rays, hot pixels, a zero bias, and a random distribution of circular masks. 
The parameters of these observational effects, e.g., noise levels, etc., span a range of realistic values (see Methods section for details). Since these effects are randomly generated at each training step, we never encounter two identical realizations of the training data. Combined with the large size of the training set, this significantly mitigates the risk of overfitting. Masks added during training are included to allow for the possibility of masking undesired artifacts in real data that the networks have not been trained on, e.g., extremely bright cosmic rays and ghosts. Since these masks are allowed to partially cover up to $25\%$ of the arcs' flux, they also render the networks insensitive to incomplete data. To further increase our accuracy, we combine the predictions in a final trainable layer. 

Our validation and test sets are both produced with the same pipeline, but with different random seeds and using background galaxy images that were not used to generate the training set (Extended Data Fig. 1). We quantify the accuracy of our predictions by calculating the interval containing 68\% of the predicted parameters from their true values. Our final $68\%$ errors from the combined network on the lensing parameters are $0.02\arcsec$, $0.04$, $0.04$, $0.04\arcsec$, and $0.04\arcsec$ for the Einstein radius, $x$, and $y$-components of ellipticity, and the $x$ and $y$-coordinates of the center of the lens, respectively. 
These errors are comparable to typical uncertainties on the estimated parameters from lens modeling with maximum likelihood methods for images with similar quality and noise levels\cite{SLACS,sonnenfeld}.
Figure 1 shows the estimated parameter values of the combined network as a function of their true values. The gray points show the parameters of $10,000$ test samples. The blue shaded regions show the $68$ and $95\%$ inclusion intervals. Table 1 summarizes the 68\% errors of the individual and combined networks.

In addition to the multiply-lensed images of background sources, optical data often include light contamination from lensing galaxies. Prior to lens modeling, this light is commonly removed in a preprocessing step by fitting a model, e.g., S\'{e}rsic, to the light distribution of the lens while masking the lensed arcs, requiring an additional supervised optimization procedure\cite{sonnenfeld}. Moreover, lensing galaxies often include complex structures not captured by simple parametric models, resulting in significant residuals. 

To fully automate the process of parameter estimation, we propose to use Independent Component Analysis (ICA) to separate the light profiles of the lens and the source arcs using multi-wavelength data. ICA is a method for separating an additive mixture of independent signals into their sub-components. In this context, the morphologies of the background and foreground galaxies are statistically independent. The color difference between these galaxies (both intrinsic and due to redshifting) results in different linear combinations of their light in different filters. Therefore, the separation of two components from two filters using ICA can help remove the lens light from the background arcs. 
We note that intrinsic color variations in the source and lens galaxies and the effects of blurring with different PSFs can result in imperfect ICA separation. However, we find that these issues only have a small effect on the resulting images and do not impact the accuracy of the recovered lensing parameters.

We demonstrate the application of this pipeline on real data by analyzing the \textit{Hubble Space Telescope} (\textit{HST}) images of SL2S survey lenses and comparing our estimates with previously published values\cite{sonnenfeld}. From this sample, we select all grade A lenses with at least two Wide Field Camera 3 (WFC3) filters and a minimum signal-to-noise ratio (SNR) of 10 per pixel. The brightest cosmic rays and the innermost region of the lens in the resulting subsample of nine systems are zeroed with circular masks that were included in the training of the networks. ICA is then applied to the images (Extended Data Fig. 2). Figure 2 shows the resulting separated arcs.
These images are then directly fed into our networks. The color markers in Figure 1 show the obtained parameters for all nine lenses. The values on the $x$-axis and their uncertainties are taken from a previously published work\cite{sonnenfeld}. The $95\%$ uncertainties of published values are consistent with our estimated values within the error of the network, an accuracy sufficient for most studies\cite{Sonnenfeld:15}.

The only manual preprocessing step in this analysis was the trivial masking of the brightest cosmic rays before the application of ICA.
Note that at the time of parameter estimation, these networks do not require the parameters of observational effects, e.g., the PSF, to be specified. Although only five parameters have been predicted here, estimating these same parameters from maximum likelihood methods requires the inclusion of hundreds to thousands of other nuisance parameters in the model, describing the morphologies of background sources\cite{warren03,suyu06}, significantly increasing the complexity and the computational cost of the parameter search. 
Moreover, these networks could be modified and trained to predict the morphologies of background sources in addition to a larger number of parameters for more complex density structures with negligible additional computational cost at evaluation time. Assuming that the typical analysis of a lens with the current profile may take a few days to complete, this method offers about seven orders of magnitude improvement in speed while automating parameter estimation. This improvement could be even greater when analyzing more complex density structures.

We have also trained Inception-v4 to estimate lensing parameters in the presence of lens light, without the application of ICA. The $68\%$ errors of this network are $0.07\arcsec$, $0.1$, $0.1$, $0.04\arcsec$, and $0.04\arcsec$, higher than those of lens-removed images, but showing significant promise for further development, especially if color information is provided. 

Although in their current forms our networks only predict global lens parameter solutions, it is in principle possible to use neural networks for parameter uncertainty estimation and posterior mapping using dropout\cite{dropout} or Bayesian networks\cite{bernouli}. Using these methods, these networks will be able to provide insight into parameter degeneracies, including multi-modal posteriors, an important issue for systems consisting of a large number of degenerate lenses, e.g., in clusters\cite{frontierfields}.

Neural networks provide a fast alternative to maximum likelihood methods commonly used to estimate parameters of interest in astrophysics from imaging data. Their effectiveness for image analysis makes them a powerful tool for applications beyond lensing studies, including stellar mass estimation from multi-wavelength data and dust temperature measurements. By streamlining these tasks, these networks can extend the reach of fast parameter estimation to general users.

\clearpage

\begin{addendum}

\item[Acknowledgements] 
We thank Ryan Keisler, Gilbert Holder, Roger Blandford, Risa Wechsler, and Warren Morningstar for useful discussions and comments on the manuscript. We also thank Gabriel P. Maher and Anjan Dwaraknath for comments about neural networks, leading to improved performance. 
 We thank Stanford Research Computing Center and their staff for providing computational resources (Sherlock Cluster) and support.
Support for this work was provided by NASA through Hubble Fellowship grant HST-HF2-51358.001-A awarded by the Space Telescope Science Institute, which is operated by the Association of Universities for Research in Astronomy, Inc., for NASA, under contract NAS 5-26555.
P.J.M. acknowledges support from the U.S. Department of Energy under contract number DE-AC02-76SF00515.

\item[Author Contributions] 
Y.D.H. and L.P.L. have contributed equally to all aspects of this project, including design and implementation of the networks and the text of the article. L.P.L. developed the use of ICA for lens removal. 
P.J.M. has contributed to various aspects of this project, including the choice of tensor libraries and tests on real and simulated data.

\item[Author Information] 
Reprints and permissions information is available at www.nature.com/reprints. The authors declare that they have no competing financial interests. Readers are welcome to comment on the online version of the paper. Correspondence and requests for materials should be addressed to Y.D.H.~(email: hezaveh@stanford.edu) and L.P.L. (email: lplevass@stanford.edu).

\end{addendum}

\clearpage

\begin{table}[]
\centering
\label{my-label}
\begin{tabular}{llllll}
Network               & $\theta_{E}$ & $\epsilon_{x}$ & $\epsilon_{y}$ & $x$ & $y$ \\ \hline
Network 1 (Inception) & 0.03 & 0.04 & 0.05 & 0.06 & 0.06 \\
Network 2 (AlexNet)   & 0.03 & 0.04 & 0.04 & 0.05 & 0.06 \\
Network 3 (Overfeat)  & 0.04 & 0.05 & 0.05 & 0.06 & 0.06 \\
Network 4             & 0.03 & 0.05 & 0.06 & 0.05 & 0.05 \\ \hline
Combined Network      & 0.02 & 0.04 & 0.04 & 0.04 & 0.04
\end{tabular}
\caption{{\bf Errors of the individual and combined networks.} The columns present the $68\%$ errors for the Einstein radius, $\theta_{E}$, the two components of complex ellipticity ($\epsilon_{x}$, $\epsilon_{y}$)  and the coordinates of the lensing galaxy ($x$, $y$). The angular parameters ($\theta_{E}$, $x$, and $y$) are given in units of arc-seconds.}
\end{table}

\clearpage

\begin{figure}[h] 
\begin{center}
\begin{tabular}{c} 
\includegraphics[width=17cm, trim=2.0cm 0.2cm 4.0cm 4.0cm, clip=true]{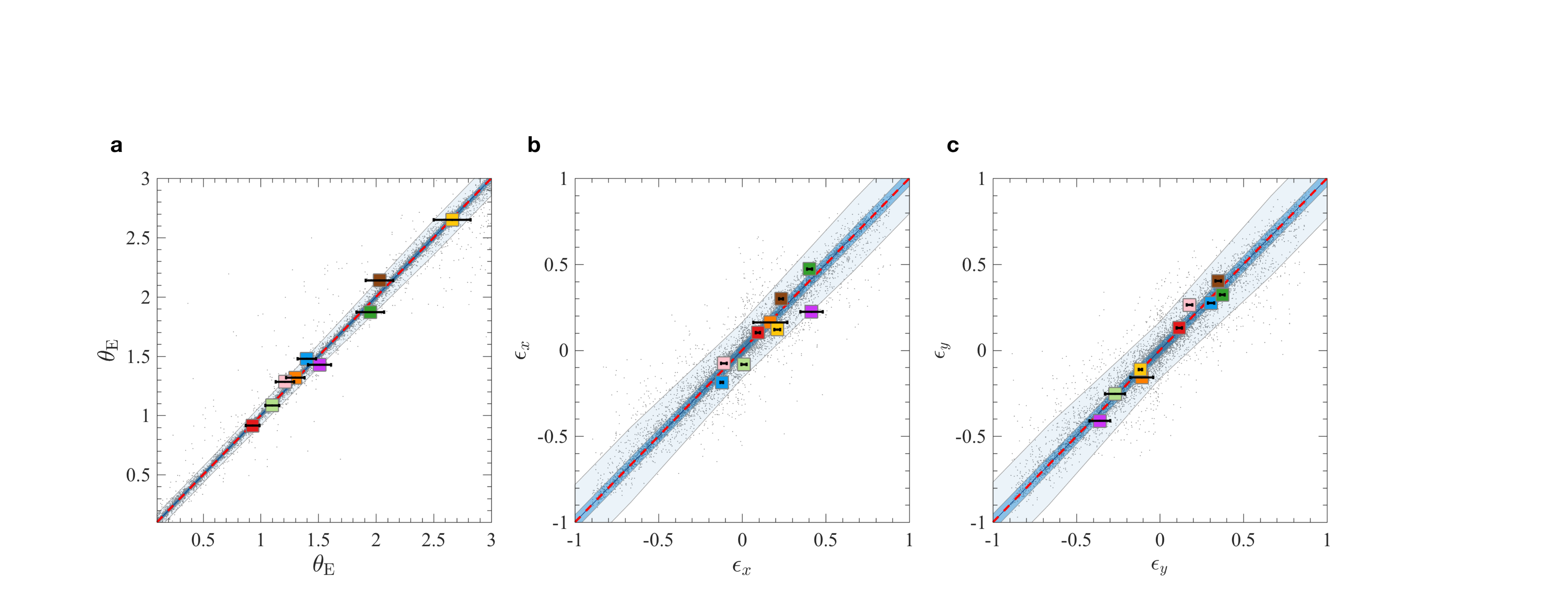} \\
\end{tabular}
\end{center}
\vspace{-2.2em}
\caption{Comparison of parameters estimated using neural networks (on the $y$-axis) with their true values ($x$-axis). From left to right, the panels correspond to the Einstein radius and the $x-$ and $y-$ components of complex ellipticity.
The shaded blue areas represent the 68, and 95\% intervals of the recovered parameters on a test set that the network has not been trained on. The small gray dots show the parameters of all 10,000 test samples. The colored data points and their error bars (95\% confidence) correspond to real \textit{HST} images of gravitational lenses in SL2S sample, with the true parameters set to previously published values\cite{sonnenfeld}.}
\label{fig:fig1}
\end{figure}

\begin{figure}[h] 
\begin{center}
\begin{tabular}{c} 
\includegraphics[width=15cm, trim=0.1cm 0.1cm 0.1cm 0.1cm, clip=true]{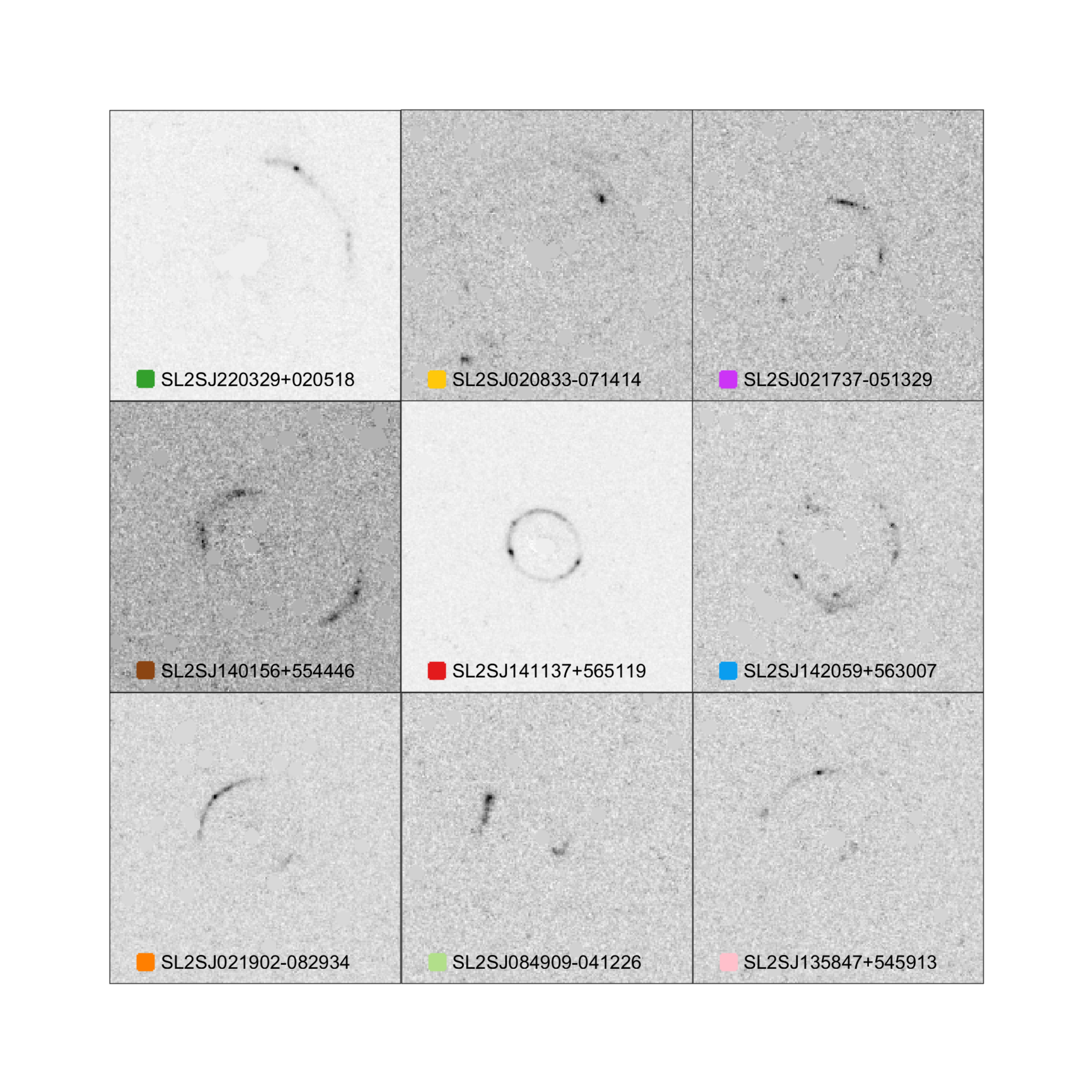} \\
\end{tabular}
\end{center}
\vspace{-2.2em}
\caption{\textit{Hubble Space Telescope} images of nine strongly lensed galaxies from the SL2S survey. These images are used to demonstrate the performance of the network on real data. The light of the lensing galaxies have been removed using Independent Component Analysis of two filters and circular masks with a radius of $0.2\arcsec$ have been applied to bright cosmic rays and the lens center. Each panel contains the object name in addition to the data marker used to show its parameters in Figure 1.}
  \label{fig:fig1}
\end{figure}

\clearpage

\appendix
\noindent{\bf METHODS}\\
A feed-forward neural network is a collection of processing units (referred to as neurons) designed to  identify underlying relationships in input data. Neurons are organized in layers; each layer's output being the input of the next layer. The output of any individual neuron can be written as $f(\bf w x)$ where ${\bf x}$ is a vector input to the neuron (for example, pixel values of an image), ${\bf w}$ is a matrix of weights (determined through training) and $f$ is a non-linear function referred to as an activation function. A network processes a given input ${\bf X}$ and maps it to an output ${\bf Y}=F_{\bf w}({\bf X})$.
A set of training data \{${{\bf X}_{train},{\bf Y}_{train}}$\} are used to determine the values of the weight matrices for this mapping. This is performed by minimizing the deviation between the network's predictions $F_{\bf w}({\bf X}_{train})$ and the true values ${\bf Y}_{train}$ for the examples in the training set by optimizing a cost function with respect to the weight matrices. The universal approximation theorem states that, under mild assumptions, a two-layer neural network can approximate any continuous function with arbitrary accuracy with a finite number of neurons\cite{universaltheorem}. However, in practice, such a simple network would require a large number of neurons and would be difficult to train. More layers can allow networks to identify higher order complex correlations with fewer parameters (weights) and to be more easily trained. In convolutional neural networks, the weights of a layer are organized in multiple two-dimensional structures, representing a set of filters. The values of neurons are then the result of the convolution of the layer's input with these filters. This allows networks to retain two- and three-dimensional structures of input data and to extract specific patterns, defined by weight filters, from them.
To perform maximum-likelihood lens modeling, given a set of lensing parameters ${\bm p}$, a simulated image, ${\bm M}=L({\bm p})$, is produced and compared to real data. Here $L$ is the operation that maps the vector of parameters to the simulated images, e.g., ray-tracing and convolution with a PSF. Our networks, on the other hand, learn the inverse of this function, mapping each image to a vector of lensing parameters ${\bm p}$. Given the flexibility of these networks to approximate complex functions and the two-dimensional structure of images, convolutional neural networks are well-suited to estimate lensing parameters directly from image data.

\noindent{\bf Simulated Datasets.}
Our datasets use a mixture of images of real and simulated background galaxies. The real galaxies comprise $60,000$ images from the Galaxy Zoo project and $8,000$ images from the GREAT3 challenge data to produce half a million lensed images. The Galaxy Zoo images are averaged over the color channels to produce a grayscale image. The simulated clumpy galaxies are composed of $1-5$ clumps. Their spatial distributions follow a randomly generated correlated Gaussian with a long axis defined as the galaxy radius. Each clump follows either a S\'{e}rsic profile with $n=1-5$, $R_{\mathrm{eff}} = 0.1-0.2\arcsec$, and $\epsilon=0-0.4$ or a Gaussian profile. 
The apparent unlensed size of all galaxies is allowed to span the range of $0.05-0.8\arcsec$. The lens Einstein radius is sampled from a flat distribution over $0.1-3.0\arcsec$. Lens ellipticity and angle are also chosen at random from a flat distribution and converted to the $x$ and $y$-components of ellipticity with a maximum ellipticity of $0.9$. The lens center coordinates cover the central $0.5\arcsec$ of the image. To ensure that the position of the sources behind the lens results in strong lensing configurations, we calculate the lensing caustics and place the sources at random inside or close to the edges of the caustics. We also ensure that the lensed images have a minimum total flux magnification of $2.0$.
The images consist of $192\times192$ pixels, with a pixel size of $0.04\arcsec$, equal to \textit{HST}-WFC3 pixels. Note that for other input pixel sizes, it is possible to interpolate the images on this grid. 
During training, at each step of a gradient descent optimization, we add realistic observational effects to the images. We first convolve the images with a Gaussian filter with a randomly chosen root mean square (rms, maximum of $0.1\arcsec$) to simulate the blurring effect of the PSF. We then convert the normalized intensity of our images to photon counts using a factor of 100-1000, and generate a Poisson realization map with a mean of these values, effectively adding Poisson noise to the images. 
We then add random Gaussian noise with either a white power spectrum or a power spectrum measured from blank fields of a sample of \textit{HST} images. The rms of the noise is chosen randomly from a flat distribution with a minimum of 1\% and a maximum of 10\% of the peak signal.
To make the networks insensitive to pixel artifacts and low-intensity cosmic rays, we generate 400,000 images containing simulated cosmic rays and hot pixels. To produce each cosmic ray, we choose a random pixel and populate the neighboring pixels sequentially, using a friends-of-friends algorithm, until the desired length is achieved. The total number of pixels for each cosmic ray and the total number of such events are also chosen at random for each map. Each artifact map may also include up to 100 hot pixels. These maps are then selected at random and added to the lensed images at the training time.
The intensity of the resulting images are normalized to 1, with a small deviation (rms of $0.05$) and a zero bias is then added to the images (Gaussian, rms$=0.05$). We then generate a number of circular masks with a radius of $0.2\arcsec$, placed at random throughout the images. Masks are allowed to cover up to $25\%$ of the image flux. The test set is made from $10,000$ GREAT3 source images that the networks have not been trained on. For testing, we only allow less than $2\%$ of the lensed flux to be masked. Extended Data Fig. 1 shows a few examples from the test set.

\noindent{\bf Training.}
The networks include a first added layer to remove the image intensity biases by filtering the maps with a $4\times4$ flat filter and subtracting the minimum of the resulting maps from the input images while excluding the masked regions from this calculation.
Due to the different nature of this problem from the classification tasks of ImageNet, we do not initialize our networks from previously trained weights on this dataset. Instead, we either use a Xavier initialization or an initialization from a fixed normal distribution and train entire networks with the Adam optimizer. The network that we designed (in addition to Inception-v4, AlexNet, and Overfeat) consists of 8 convolutional modules. Each module consists of a convolution with a large filter, followed by a $1\times1$ convolution with the same depth. Third, fifth, and the seventh modules use stride 2 convolutions to reduce the image size. The sizes of the primary kernels are 3, 5, 10, 10, 10, 10, and 3, with depths of 32, 32, 32, 32, 64, 64, 128, and 256. These are then passed to two fully connected layers with 512 and 5 (output) nodes. All layers except the last have a rectified linear activation. We calculate a Euclidean mean square loss averaged over the five parameters.  Because of the parity invariance of the components of ellipticity, we calculate this loss for both ($\epsilon_x$, $\epsilon_y$) and ($-\epsilon_x$, $-\epsilon_y$), keeping the minimum of the two loss values. 
The large size of our training data and the application of randomly generated observational effects for each encounter of the training examples mitigate the risks of overfitting. We do not use dropout layers and notice that even with long training the average cost value for the training data does not exceed that of the validation set. About one day of training on a single graphics processing unit can result in modest accuracy. Improving this accuracy to that quoted in the main text requires an additional few days of training.
We note that neural networks can have a large number of local minima, with some having poorer prediction performance than others. It may sometimes be necessary to  restart the training of networks to find a better local minimum than the current solution. However, it has been empirically observed that the highly non-convex cost function of such networks can usually be relatively easily optimized using local updates\cite{localminima}.
To average the results of the four networks, we combine their individual predictions in a final trainable layer. For this, we fix the four networks and only train the parameters of this layer alone.

\noindent{\bf SL2S data and ICA.}
We obtain the SL2S \textit{HST}-WFC3 data from the Mikulski Archive for Space Telescopes. We select the lenses classified as grade A\cite{sonnenfeld} with a minimum SNR of 10 per pixel, with at least two available WFC3 filter images. The images are visually cropped to include the lensed arcs near the center of the images. We then apply circular masks of radius $0.2\arcsec$ to the bright cosmic rays and to the brightest regions of the lensing galaxies. The images are then passed to a fast ICA code to be unmixed into two subcomponents. Extended Data Fig. 2 shows the resulting separated ICA components from the two filters.
The resulting images of the arcs are then fed into the networks. We note that true color variations in each of the galaxies and PSF differences in the two images can result in imperfect component separation. To test if such effects influence the decisions of the networks in determining lensing parameters we also subtract the light of the lensing galaxies with a S\'{e}rsic profile. We do not notice a significant difference in the estimated parameters using the two different approaches. To convert the published uncertainties of ellipticity and angle\cite{sonnenfeld} to uncertainties in $x$ and $y$-components of ellipticity, we assume an uncorrelated normal posterior for the polar parameters. 
For SL2SJ220329+020518, previous analysis excluded a central blue arc, initially thought to be a part of the background source, from  lens models, based on an additional spectroscopic analysis\cite{sonnenfeld}. We similarly masked this image prior to feeding it into our pipeline.

\noindent{\bf Tests on robustness.}
We evaluated the performance of the networks for tests convolved with a realistic (non-Gaussian) \textit{HST} PSF and found that they did not reduce the parameter accuracies significantly.
It is expected that the performance of the networks on images with lensing parameters outside the range explored in training data or with a new class of nuisance features that the network has not been trained on (e.g., a satellite track) will decrease with the strength of these effects. For example, we tested the performance of our networks, trained on data with no external shear, on new examples of lensing systems which included external shear with a maximum shear of $0.3$. The root mean square errors of the networks for Einstein radius and complex ellipticity rise steadily for increasing shear while the lens center is robustly predicted at all shear values. 
The rms errors for ellipticity rise from $0.11$ for systems with shear less than $0.01$ to $0.23$ for lenses with shear more than $0.15$. We note, however, that these errors could be significantly reduced by training the networks on systems which include external shear. Approximate Bayesian neural networks, which produce the posterior of the parameters for each example, can capture the uncertainties of the networks in their predictions for test examples outside the training space\cite{bernouli}. 

\noindent{\bf Code availability.}
The code used to simulate the training, validation, and test sets, the code to estimate the lensing parameters, and the trained network weights are freely available for download at https://github.com/yasharhezaveh/Ensai.

\noindent{\bf Data availability.} The \textit{HST} data used to test the performance of the networks are publicly available for download from the Mikulski Archive for Space Telescopes (https://archive.stsci.edu/hst/search.php). The unlensed images of background galaxies used to simulate the training, validation and test sets can be obtained from the GREAT3 website (http://great3challenge.info) and the Galaxy Zoo machine learning challenge web page (https://www.kaggle.com/c/galaxy-zoo-the-galaxy-challenge).

\clearpage

\renewcommand{\thefigure}{S.\arabic{figure}}

\begin{figure*}[ht]
\centering
\includegraphics[width=15.cm,angle=0]{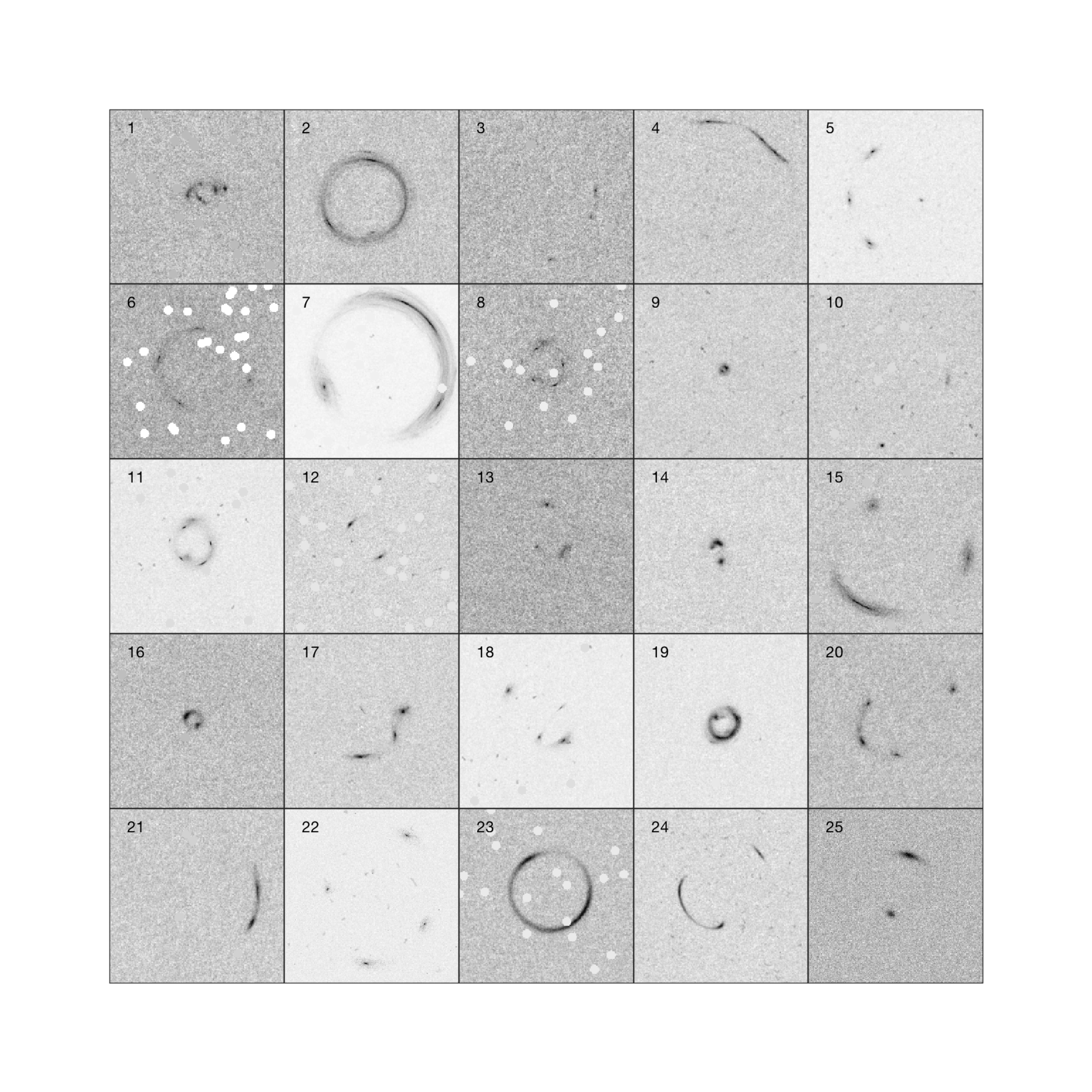}
\caption{A selection of the test samples used to evaluate the performance of the network. These examples are chosen to illustrate the variations of different effects, including cosmic rays (e.g., panels 11 and 12), masks (e.g., panels 6 and 23), Einstein radii (e.g., panels 7 and 9), noise levels and PSF blurring strengths, and a mixture of lensing image configurations including some unfavorable morphologies (e.g., panels 10 and 21).}
\label{fig:S1}
\end{figure*}

\clearpage

\begin{figure*}[ht]
\centering
\includegraphics[width=14.cm,trim=1.0cm 1.0cm 1.0cm 1.0cm, angle=0]{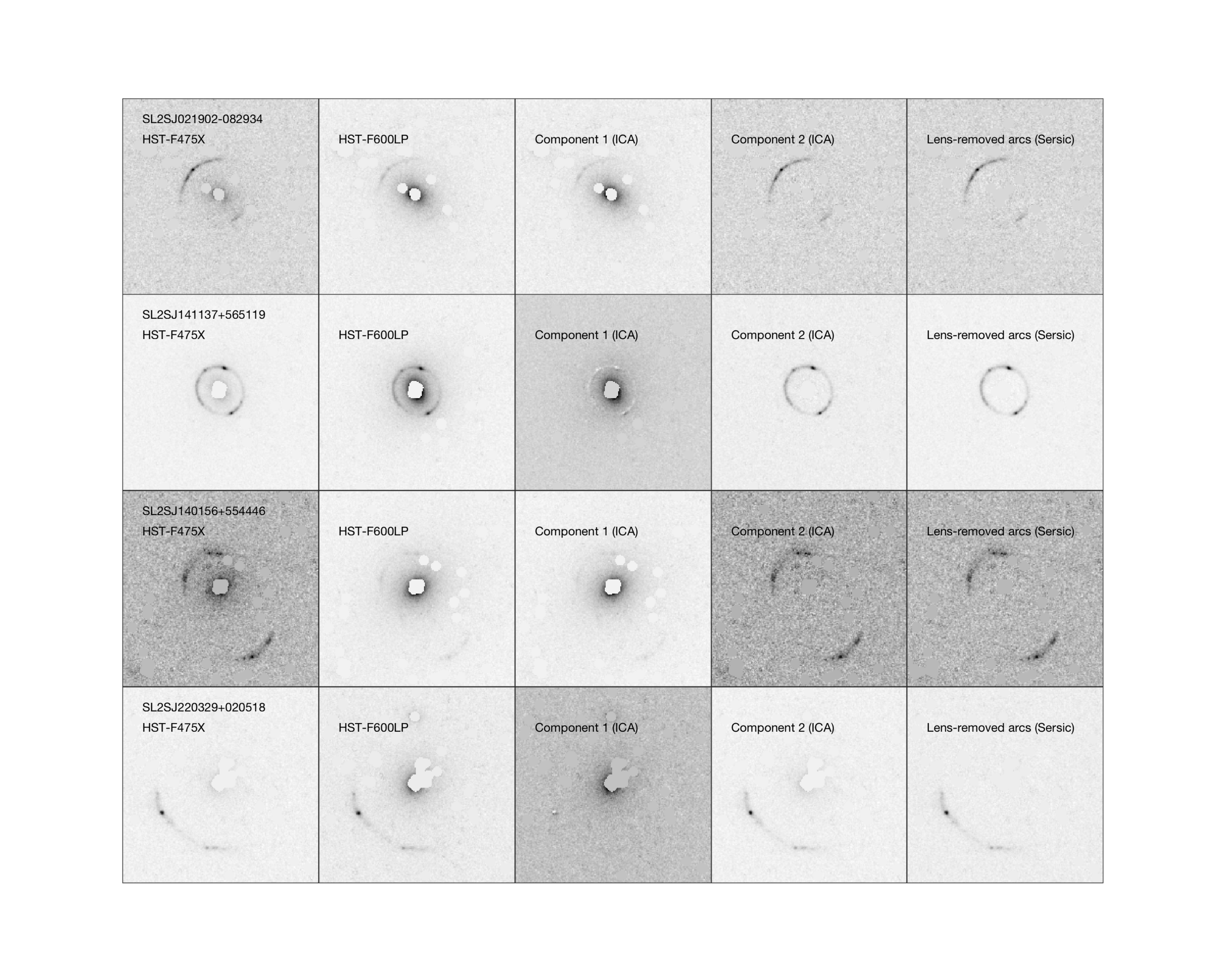}
\caption{Examples of the inputs and outputs of the ICA. For each row the first two panels show the \textit{HST} images in F475X, and F600LP filters. The third and fourth columns show the outputs of the ICA. For comparison, lens-removed arcs using a S\'{e}rsic model are shown in the last column. Cosmic rays and the brightest central parts of the lensing galaxies have been masked.}
\label{fig:S2}
\end{figure*}

\end{document}